\documentclass[runningheads]{llncs}

\usepackage{eccv}
\usepackage{eccvabbrv}
\usepackage{graphicx}
\usepackage{booktabs}
\usepackage[accsupp]{axessibility}  
\usepackage[breaklinks,colorlinks,citecolor=eccvblue]{hyperref}
\usepackage{orcidlink}
\usepackage{xspace}
\usepackage{booktabs}
\usepackage[english]{babel}
\usepackage{caption}
\usepackage{subcaption}
\usepackage{wrapfig}
\usepackage{float}
\usepackage{amsmath}
\usepackage{a4wide}

\begin{document}

\title{Dynamic Social Networks in Dairy Cows} 
\author{Emil Grosfilley\textsuperscript{$*$}, Yujie Mu\textsuperscript{$*$}, Dap De Bruijckere\textsuperscript{$*$}}
\authorrunning{Emil Grosfilley, Yujie Mu, Dap De Bruijckere}
\institute{Independent Researcher(s)\\}

\maketitle
\def\thefootnote{$*$}\footnotetext{Equal contribution.}\def\thefootnote{\arabic{footnote}}

\begin{abstract}
Social relations have been shown to impact individual and group success in farm animal populations. Fundamental to addressing these relationships is an understanding of the social network structure resulting from the co-habitation and co-movement of relationships between individuals in a group. Here, we investigate the social network of a group of around 210 lactating dairy cows on a dutch farm during a 14 days period. A positioning system called \emph{Cowview} \cite{Cowview} collected positional data for the whole period. We make the assumption that spatial proximity can be used as a proxy for social interaction. The data is processed to get adjacency matrices. Then social networks are identified based on these matrices. Community detection techniques are applied to the networks. We measure metrics of different dimensions to test community structure, centralization, and similarity of network structure over time. Our study show that there is no evidence that cows are subdivided into stable social communities when looking at interaction in the whole barn. We, however, notice relatively clear communities when dividing the barn into areas with different activities.
The social network is characterized by significant centralization, low connectivity, and a hierarchy.
\end{abstract}

\section{Introduction}
As the dairy industry becomes more aware of the impact of the social environment on animal welfare and production, there is a growing need for information on the optimal size, stocking densities and composition of cow management herds. In a stable social group, many cows form preferential social bonds, which may vary between activities such as feeding or social grooming. Interestingly, social grooming is related to production \cite{gygax2010socio} and it has been positively correlated with milk production and weight gain in past studies. Cows' social preferences are also reflected in their spatial proximity to others in the group, so the ability to maintain suitable inter-individual spaces is important for cattle \cite{SATO19913}\cite{boe2003grouping}. Numerous studies have demonstrated negative effects of restructuring on welfare and productivity, including reductions in milk production, feed intake, rumination and bedtime, and increased aggression among cows \cite{HASEGAWA199715}\cite{HULTGREN2009255}.

Agricultural systems need to improve in order to meet future challenges of an increasing human population and at the same time ensure good animal welfare and a minimal ecological footprint. Thus, a project called \emph{'Precision livestock breeding - improving both health and production in dairy cattle'} \cite{sluproject} is in progress. This project focuses on how to improve animal welfare and health in dairy cattle production using sensor technology, with the potential to minimize antibiotic use by improving tools for disease control and management. 

Our work is part of the above project and is to investigate how consistent social networks are between days and how they develop over time. Understanding the dynamics of cows' social networks can allow them to make better decisions on how to split a group in the case of disease transmission and improve the opportunity of cows to preserve meaningful social contacts to enhance their well-being, increase milk production, and growth. We explore these social network dynamics based on social network analysis (SNA) and studies on cows' social behaviour \cite{CHEN}. 

Social network analysis (SNA) has been developed to quantitatively measure and analyse the structure of groups and patterns caused by dyadic social interactions \cite{croft2008exploring}. A network is made up of nodes (individuals: cows in this case) and edges (interactions: association time in this case). We can calculate statistics for individuals in the network, such as 'degree' (number of edges for a given node) and 'betweenness centrality' (number of shortest paths between pairs of individuals that pass through a particular individual) \cite{krause2009animal}. These methods allow us to study non-random patterns of association and detect differences in group structure that may be linked to individual attributes. SNA is becoming more popular in the field of animal behaviour.

In this report, we describe the problem we try to deal with in detail in Section 3. According to previous studies  \cite{CHEN}\cite{Farine2015ProximityAA}, we quantified the social network structure of a group of lactating dairy cows based on the interaction time in specific areas which comes from the data collected by \emph{Cowview}. We analyzed the centralization, edge distribution, eigenvalues, etc., to study the hierarchical features of dairy cow social networks. We used different community detection methods and community measurements to study the features of dairy cow communities. By comparing the features of the 14-day networks as well as the communities in them, we studied how they change over time. The relevant methods are shown in Section 4, while the results are shown in Section 5. In Section 6, we draw some conclusions. We also describe some potential improvements in Section 7.

\section{Problem Description}
This project focuses on the dynamics of the social network of dairy cows.  
In more detail, what the cows' social network and communities look like, how the cows' social network and community change over a two-week period, and when new cows join, how they affect the cows' social network and community.

In order to solve the above problems, we select different attributes of social networks and analyze them on different "dimensions"; monadic measurement, dyadic measurements, and polyadic measurements. Monadic means that each measurement uses one characteristic, one example is how many edges a cow has. Dyadic measurements use two characteristics which can be how central a cow, and its neighbouring cows are to get the measurement. Polyadic means that more than two characteristics are used for each measurement. We also study a variety of popular community detection algorithms based on the behavioral habits of cows \cite{CHEN}. The best performing method are selected and the communities it detects are analyzed. Both the attributes of social networks and communities over a single day and compare the changes in their attributes over a two-week period are studied.  In addition, we also perform random simulation experiments to demonstrate that the social network and community of cows are very different from random simulations.

Combining the analysis of these different metrics in different dimensions gives us a more complete view of what social interactions look like, while the comparison of the results over 14 days show how the social network develops in time.

\section{Converting positional data to social networks}
\subsection{Social network}
A network consists of nodes, in our case the cows, and edges, which connect the nodes; in our case, these represent a social contact. Social network analysis is a very powerful tool to analyse social structure. The analytical techniques and algorithms are well developed and there exists a large number of functions to execute them in python, for example, the package {\tt networkX}.

\subsection{Convert positional data to time adjacency matrices}
\begin{figure}[h!]
    \centering
    \includegraphics[width=0.8\textwidth]{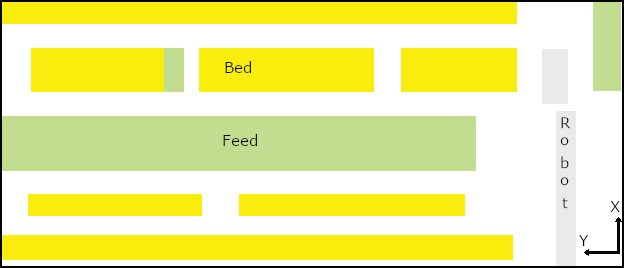}
    \caption{The layout of the barn showing the beds, feeding and robot areas. White represents the general area.}
    \label{fig:barn}
\end{figure}

At our disposition, we have positional data of around 200 - 220 dairy cows over a duration of two weeks, coming from a farm in the Netherlands, (layout in Figure \ref{fig:barn}). The number of cows can change from day to day due to some new cows arriving and some already inhabiting cows leaving the barn during the data collection period.

The data is collected in the enclosed barn using a real-time location system (RTLS) developed by \emph{GEA systems} called \emph{Cowview} \cite{Cowview}, consisting of multiple receivers located around the barn and individual tags that were attached with a collar around every cow's neck. There are also some tags that are not attached to a cow.

The RTLS calculates the position by combining multiple signals picked up from the receivers and triangulating the position of the tags, which leads to a positional accuracy of 50 cm, and time accuracy in the order of milliseconds. The data consisted of one CSV file for each day of collected data, where each row represented an occurrence of data logging. A row of data contained the tag's id, the x and y position in the barn measured in centimeters, the time of logging, and additional information we did not use.

We converted the positional data into an time adjacency matrix (TAM).

A TAM is an $N \times N$ square matrix, where $N$ is the number of cows, logging the time every pair of cows have been adjacent. For example, the interaction time for cow one and cow two would be logged in index $[1,2]$ and $[2,1]$.

To be counted as adjacent, two cows need to be closer than an arbitrary limit. We chose this limit to be 150cm, as it is the approximate length of a cow.

As the positional data of each cow is logged around once a second, and there are around 210 cows in the barn in a day, the amount of data is quite large, consisting of about 12 million data points per day. 
Thus we need to compromise between the time resolution and the time to compute the TAM. 

A 10s time resolution turned out to be a good compromise. It led to acceptable run-times and, most importantly, only changed 60 cells in the binary adjacency matrix when compared to 1s resolution. The matrix contains around 44000 cells; when using 1s time-steps, the number of edges is around 3000. The 10s resolution, when compared with the 1s resolution, preserves about $ 98\% $ of the information. 10s is also the time resolution used by Chen \cite{CHEN}.

The TAM was thus computed by looking at the distance between each pair of cows, where the latest recorded position was used, every ten seconds. If the distance between a pair was smaller than the 150cm limit, the cell corresponding to the pair was increased by ten.

Using the data for one day, we are thus left with an TAM in which each cell corresponds to the time that pair has spent in proximity during that day, measured in seconds.

As mentioned before, there are some tags that were not placed on cows; we thus need to filter our tags to only have cows in our TAM. This was done by removing all data corresponding to tags that did not move more than 500cm in the y-direction (Figure \ref{fig:barn}). This can be done as all cows present need to move to the milking robot situated at the bottom of the barn and are thus guaranteed to move more than 500cm in the y-direction during one day.

In Chen's article \cite{CHEN} and Farine's article \cite{Farine2015ProximityAA}, there is an argument that social networks can only be seen when looking at interaction in particular areas, for example, that cows have a preference with whom they eat close to. We thus also made area-specific TAM.

Here the barn was divided into four discontinuous areas (Figure \ref{fig:barn}), one 'Bed' area where there are beds, one 'Robot' area where the milking robot is located, one 'Feed' area where the cows are fed, and finally a 'general'  area, which encompasses all interaction which is not in one of the first three areas.

The same data was used for the TAM. The output was, however, divided into four TAM, one for each area. The adjacency of a pair of cows was only recorded in one of the matrices, where the specific areas take precedent over the general area. The specific areas were far enough apart to never clash.

We thus have turned the positional data into TAM, both daily TAM and daily area-specific TAM. These can further be used to look at the social networks of dairy cows.

\subsection{Binary and weighted adjacency matrices}
Using the TAM concept, we can use both a Binary adjacency matrix (BAM) and a weighted adjacency matrix (WAM). Both are a $N \times N$ square matrices, with zeros meaning the pair of cows does not achieve the minimum required time in proximity. A connection between cows is represented as 'one' in the BAM and as a number between zero and one depending on its weight in time int the WAM.

The minimum required time for proximity is an important concern to consider when modeling social networks. If it is too large, some meaningful connections might be filtered out, while a too short time would mean an overcrowded and possibly meaningless network. In \cite{ROCHA2020104921} there is an argument that the minimum time is not changing the network qualitatively within reasonable boundaries. However, they have a sizeable quantitative effect on the network. It is thus better to choose a sizeable minimum time so that the analysis of the network becomes easier while keeping it small enough not to lose too many meaningful interactions. We decided to go with a 30 minutes limit. The formula are expressed as 
\begin{equation}
weight_{i,j} =
\Bigg\{
\begin{matrix}
  \frac{t_{i,j}-t_{min}}{t_{max}-t_{min}} & if \quad t_{i,j}>=t_{min} & \\ 
  \\
 0 & otherwise;
\end{matrix}
\label{eq:weight}
\end{equation}

$$t_{max}=max(t_{i,j}).$$
 The time at a cell ($t_{i,j}$) needed to be more or equal than the minimum time ($t_{min}$) required to be a connection, only then  eq. (\ref{eq:weight}) will be used. $t_{max}$ is the maximum time two cows have spent together that day, which should be bigger than $t_{min}$ because all the connections will have a time equal or between $t_{min}$ and $t_{max}$. This creates a matrix with elements with values between zero and one since all times below the minimum time requirement are set to zero, with higher numbers meaning they spend more time together. The advantage of this is that you can see the percentage based on the maximum time that two cows have spent together. A WAM gives the network an extra dimension by giving a weight to the edges based on the cows' time together. There are some other suggestions on how to obtain the weights. Weights are expressed as
 \begin{equation}
 weight_{i,j}=\frac{t_{i,j}-t_{min}}{t_{max}}.
 \label{wrong1}
 \end{equation}
 In eq. (\ref{wrong1}) we remove the possibility that there is an error if $t_{max} = t_{min}$ at the cost of having the weights be between zero and $1-\frac{t_{min}}{t_{max}}$ instead of zero and one. The reason we chose not to use this one is because if $t_{max} = t_{min}$ then you would not have any connections at all in the network. In this case, weights are given by
 \begin{equation}
 weight_{i,j}=\frac{t_{i,j}}{min(sum_{t_i},sum_{t_j})}.
 \label{swr1}
 \end{equation}
 $sum_{t_i}$ means the sum of all the time that cow $t_i$ has spent with other cows. The advantage of using eq. (\ref{swr1}) is that you are giving it a weight based on the total time that the cow spends with other cows, but that also means that cows with fewer connections are more likely to get higher weights. We chose to use eq. (\ref{eq:weight}). 
\subsection{Converting matrices to graphs}
Many concepts of social networks come from graph theory.
The social network model is a graph composed of nodes (cows) and edges (social relations).
Since the social relationship between the two cows is the same, the social network of cows can be represented by an undirected graph using undirected edges.
Therefore, the visualization and analysis of social networks can be performed by using graph analysis tools.
In this project, the social network is established through the {\tt NetworkX} package in {\tt python}.
The graph is obtained from one of the adjacency matrices. Because it is an undirected graph, only the elements in the upper triangular part of the matrix need to be used.

For a binary adjacency matrix, an element with a value of 0 in the matrix indicates that there is no edge between two nodes, and an element with a value of 1 in the matrix shows that there is an edge between the two nodes. The weights of these edges are all 1; that is, the resulting graph is an undirected graph without weights.

For a weighted adjacency matrix, an element with a value of 0 in the matrix indicates that there is no edge between two nodes, and a non-zero element in the matrix shows that there is an edge between two nodes. The weight of the edge is the corresponding non-zero element value. The weights of the edges range from 0 to 1. The resulting graph is an undirected graph with weights.
\subsection{Community detection}
One possible way to analyze the dynamic social network of cows is to detect the communities in the network over several days and investigate the development or changes of these communities during said days. We thus applied several community detection methods to the graphs obtained from adjacency matrices.

\subsubsection{Community detection methods on unweighted and weighted graphs}
The following community detection algorithms were applied to the unweighted graphs of the social networks in this project.

The \emph{Girvan-Newman(GN)} algorithm \cite{PhysRevE.69.026113} is a classic community detection method, which can divide a graph into multiple non-overlapping communities. The GN algorithm is a typical divisive algorithm. In detail, the edge betweenness centrality(EBC) \cite{PhysRevE.69.026113} can be defined as the number of shortest paths through an edge in the network. Each edge is given an EBC score depending on the shortest path between all nodes in the graph. The communities in the GN algorithm are found by iteratively removing the edges of the graph based on the EBC values of all edges in the graph. The edge with the largest betweenness centrality is removed first.

The limitation of the GN algorithm is that the number of communities needs to be specified in advance. In other words, the division level of the graph needs to be determined in advance. This requires a lot of prior knowledge or a validation data set with true labels for comparison. It is otherwise impossible to judge whether the communities obtained are reasonable.

The \emph{Louvain} algorithm \cite{traag2019louvain} is a community discovery algorithm based on graph data. The optimization goal of the algorithm is to maximize modularity.
The communities generated by the Louvain algorithm are also non-overlapping.
Compared with GN, the number of communities generated by the Louvain algorithm does not need to be determined manually.

Considering that the number of clusters of cows cannot be determined in advance, the Louvain algorithm is very suitable for community detection of cows.

Additionally, cows might belong to multiple communities, which means the communities may be overlapping.

The \emph{clique percolation method(CPM)} \cite{derenyi2005clique} is an algorithm for detecting overlapping communities.
In this algorithm, the community is considered as a set of fully connected subgraphs with shared nodes, and the community structure in the network is identified by looking for maximal cliques and processing them through certain rules.

Community detection methods are also applied on weighted graphs generated from weighted adjacency matrices. The results are compared with those of unweighted graphs to see if weighted graphs could detect more stable communities. Only the Louvain algorithm is applicable to the weighted graphs, so the detected communities, in this case, are also non-overlapping.

\subsubsection{Community detection on different areas}
According to Farine's article \cite{Farine2015ProximityAA}, animals are likely to have different communities in different activities. Therefore, we build networks from both the whole barn area and four specific areas, as we mentioned in Section 4.2. Relevant characteristics of communities are also analyzed.

\section{Results}
The results can be divided into three levels of metrics: Monadic metrics, which are representations of the characteristics of individuals. Dyadic metrics, which represent characteristics of pairs, can also be seen as the characteristics of the edges or interactions. Polyadic metrics look at the features of three or more individuals. Characteristics of social networks and communities are also considered polyadic metrics. All the results in the section are rounded to three decimal places.

\subsection{Monadic measurements}

\subsubsection{Centrality}
 \begin{figure}[ht!]
     \centering
     \begin{subfigure}[b]{0.49\textwidth}
         \centering
         \includegraphics[width=\textwidth]{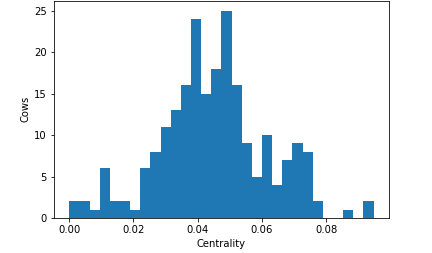}
         \caption{Average over 14 days}
         \label{fig:cent_avr}
     \end{subfigure}
     \hfill
     \begin{subfigure}[b]{0.49\textwidth}
         \centering
         \includegraphics[width=\textwidth]{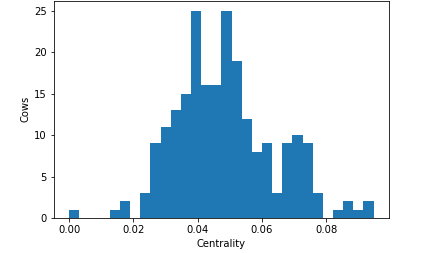}
         \caption{Average over days cows are present}
     \end{subfigure}
        \caption{The average degree centrality for all cows.}
     \label{fig:centrality}
\end{figure}

Degree centrality is a measure of how connected a node is to other nodes. In our case, the centrality of a cow can be interpreted as the risk it poses in disease contamination and how important that cow is in the social structure. It is calculated by looking at how many connections or edges a cow or node has compared to all possible nodes it could have had, which is $N-1$, where N is the number of cows for the given day. Having $N-1$ connection would be represented as having a centrality of 1. The centrality is then averaged over the period. In Figure \ref{fig:centrality} we see that there are some cows that are significantly more central than the average, as well as some cows that are significantly less central than the average.

\subsubsection{Dynamics of the Edge distribution}

\begin{figure}[h!]
     \centering
     \begin{subfigure}[t]{0.47\textwidth}
         \centering
         \includegraphics[width=\textwidth]{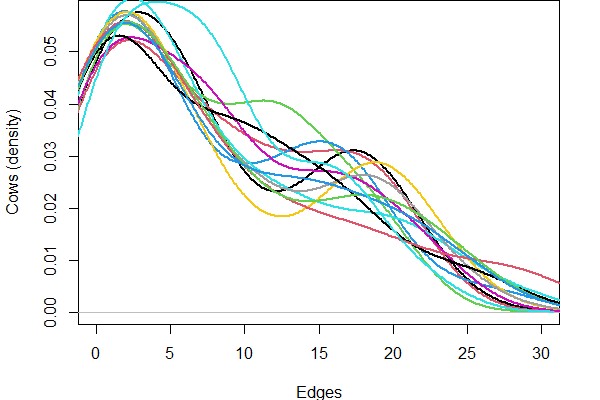}
         \caption{Density plot of the edge distributions for all days, overlapping.}
         \label{fig:Edgmult}
     \end{subfigure}
     \begin{subfigure}[t]{0.47\textwidth}
         \centering
         \includegraphics[width=\textwidth]{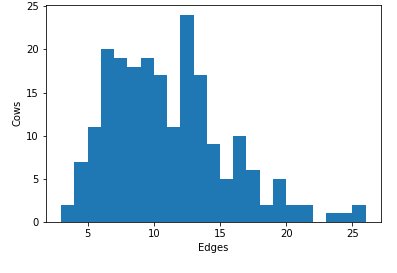}
         \caption{Histogram for Day 9}
         \label{fig:Edgesing}
     \end{subfigure}
     \label{fig:Edgedist}
\end{figure}

Another monadic measurement is the Edge distribution. This is very similar to Centrality and counts the number of edges each node has for a given day. Unlike Centrality, it does, however, not scale the result by all possible connections. Edge distribution and Centrality measure the same thing but with different scaling.

In Figure \ref{fig:Edgmult} we see the overlapping density plots of the edge distribution for every day in the 14 days period. By comparing multiple days, we can get an understanding of the dynamics of the number of connections in the network. To measure this, we use the Two-sample Kolmogorov-Smirnov (TSKS) test. The null hypothesis for this test is that the edge distributions are all from the same distribution.

For each pair of days we obtain a p-value via computing the TSKS score for each pair of days. Bonferroni's one-step correction is then applied to account for the fact that the different p-values come from the same underlying data. We use $\alpha = 0.05$, a common threshold that indicates a family-wise error rate of less than or equal to 5\%. The result is that none of the pairs reject the null hypothesis. This means that none of the edge distributions are different enough to be considered non-correlated.

To look at whether the individuals have a consistent place in the edge distribution, we can look at the top $10\%$ of individuals for each day. We notice that three cows are present in 10 or 11 of the 14 days. Significantly more than all other cows, with many cows achieving 6 or 7 days, but none between 7 and 10. This, combined with the centrality distribution from Figure \ref{fig:cent_avr}, would imply that there are three leading cows that have significantly more contact than others and that this position is relatively consistent over time.

\subsubsection{Eigenvalues}
The eigenvalues of BAM can be used to determine how similar the networks are to each other.

The largest eigenvalue, known as the spectral radius, approximately depicts the average degree distribution, which is the average amount of edges each node has. As shown in Table \ref{tab:spec_rad} it is similar to the average degree distribution of the network. 
\begin{table}[ht!]
\centering
\begin{tabular}{|c|c|c|c|}
\hline
\ Day & spectral radius & average degree distribution& spectral radius ratio   \\
\hline
1      & 11.91 &  9.76   &  1.22\\
2      & 12.51  & 10.53   & 1.19 \\
3      & 12.81   & 11.20  & 1.14 \\
4      & 11.28   & 9.49  &  1.19 \\
5      & 11.22   & 9.44  &  1.19\\
6      & 10.61   & 8.76  &  1.21\\
7      & 11.21   & 9.48  &  1.18\\
8      & 10.15   & 8.53  &  1.19\\
9      & 13.46    & 11.40 & 1.18\\
10     & 11.90  &  9.92 &   1.20\\
11     & 11.63   & 10.05  & 1.16\\
12     & 11.58   & 9.79  &  1.18\\
13     & 11.81  & 10.27  &  1.15\\
14     & 12.07  &  10.39 &  1.16\\

\hline
\end{tabular}
\caption{ The spectral radius for each day. Spectral radius ratio is the ratio between the spectral radius and the average degree distribution. }
\label{tab:spec_rad}
\end{table}

\subsection{Dyadic measurements}
\subsubsection{Eigenvector centrality}
 \begin{figure}[ht!]
     \centering
     \includegraphics[width=85mm]{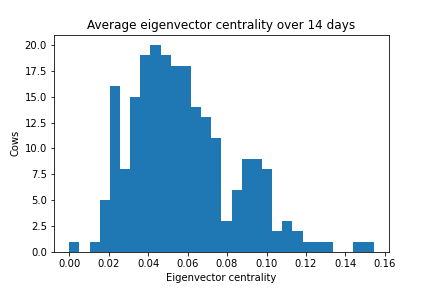}
     \caption{The average eigenvector centrality for all cows over a 14 days period. If a cow is present less than 14 days the average is taken over the days it is present.}
     \label{fig:eigcentrality}
 \end{figure}
The difference between degree centrality and eigenvector centrality is that eigenvector centrality looks at how connected a node is to other nodes and how connected those nodes are. This means that is connected to a well-connected node increases the Eigenvector centrality more than being connected to a less well-connected node. If we compare Figures \ref{fig:centrality} and \ref{fig:eigcentrality}, no major changes can be seen.
\subsubsection{Average shortest path}
The average shortest path (ASP) of a network describes how many edges on average we need to go through from one cow to another. The minimum path is one, which represents a fully connected network. The ASP can be used to see how connected the network is overall. The reason some days do not have an ASP is that the network is not fully connected, meaning that there is no path between all cows. These are one to two cows that turn out not to have connections to any other cows on that day. As is shown in Table \ref{tab:realsim_tran_1}, the ASP is consistent between the days that are connected. The diameter of the network is the maximum shortest path needed to go from one cow to any other cow in the network.

\begin{table}[h!]
\centering
\begin{tabular}{|c|c|c|}
\hline
\ Days &  Avg shortest path& diameter \\
\hline
1      & 2.605 &  5       \\
2     &  2.544 &   5\\
3      & 2.498 &   4\\
4      & -    & -\\
5     &  2.623 & 5  \\
6      & -      & -    \\
7     &  -    &-\\
8      & -    &-\\
9      & 2.465 & 4  \\
10     & 2.588  &5\\
11      &  2.557 & 4        \\
12     &   -    &-\\
13      & 2.546  & 5 \\
14      & 2.519   & 4\\
\hline
mean      & 2.549&    \\
Std      & 0.051&   \\
\hline
\end{tabular}
\caption{Average shortest path and diameter for the network each day. Days with missing data are not fully connected networks}
\label{tab:realsim_tran_1}
\end{table}

\subsection{Polyadic measurements}
\subsubsection{Transitivity}
Transitivity, also called the clustering coefficient, measures how likely nodes cluster in the social network. This shows how likely the cows in our network are to form groups. It does this by checking the number of existing triangles, three nodes connected to each other, divided by the number of possible triangles. The possible triangles are the existing ones and ones where two nodes are connected to the same node but not to each other.   
\begin{table}[ht!]
\centering
\begin{tabular}{|c|c|}
\hline
\ Days & Transitivity \\
\hline
1      & 0.0523           \\
2     & 0.0615      \\
3      &  0.0577    \\
4      & 0.0583     \\
5     &  0.0541     \\
6      & 0.0471            \\
7     & 0.0492      \\
8      & 0.0390     \\
9      & 0.0699     \\
10     & 0.0604   \\
11      & 0.0478          \\
12     & 0.0567     \\
13      & 0.0536    \\
14      & 0.0576    \\
\hline
mean      & 0.0547     \\
Std      & 0.0075     \\
\hline
\end{tabular}
\caption{The transitivity of the network each day }
\label{tab:realsim_tran}
\end{table}

By simulating 1000 networks, we can conclude if the transitivity is random or not. The simulation was done by getting the degree sequence, a list of how many edges each node has, and creating a new network with the same degree sequence. As Table \ref{tab:realsim_tran} shows, the mean of our network is significantly different from the randomly generated networks meaning that the transitivity in our network is not random.
\begin{table}[ht!]
\centering
\begin{tabular}{|c|c|c|}
\hline
\ Transitivity & Real & Simulated  \\
\hline
mean      & 0.0547 & 0.1090    \\
Std      & 0.0075 &  0.0026    \\
\hline
\end{tabular}
\caption{ The mean transitivity of our networks compared to a simulated network}
\label{tab:Avg_Tran}
\end{table}

\subsubsection{Connectivity}
The connectivity of a graph measures how many edges are needed to be removed for the graph to become disconnected. As Table \ref{tab:fiedler} shows, the connectivity fluctuates between zero and two on each day. To obtain the algebraic connectivity, the Laplacian matrix has to be used. 
The Laplacian of the network is the degree matrix minus the BAM. The lowest eigenvalue of the Laplacian is always zero.

The second smallest eigenvalue is the algebraic connectivity, also known as the Fiedler value, of the network. The algebraic connectivity is always smaller or equal to the connectivity \cite{DEABREU200753}. This shows how connected the network is by getting higher numbers meaning that the network is more connected.


\begin{table}[ht!]
\centering
\begin{tabular}{|c|c|c|}
\hline
\ Day & Fiedler value & Connectivity\\
\hline
1      & 0.887 & 1 \\
2      & 0.781 & 1\\
3      & 1.786 & 2\\
4      & 0.0   &  0    \\
5      & 0.795 & 1\\
6      & 0.0   & 0\\
7      & 0.0   & 0 \\
8      & 0.0   &  0\\
9      & 1.883 &  2\\
10      &  0.845 & 1 \\
11     & 0.914   &  1 \\
12     & 0.0     &  0\\
13      & 1.425  & 2\\
14     & 1.697   &  2\\

\hline
\end{tabular}
\caption{ The Algebraic connectivity and connectivity for each day }
\label{tab:fiedler}
\end{table}

\subsection{Results of social networks}
\subsubsection{Statistics}
Density, number of nodes, and number of edges are fundamental characteristics of a network. These metrics for our networks can be seen in Table \ref{tab:graph_stat}. 

\begin{table}[ht!]
\centering
\begin{tabular}{|c|c|c|c|}
\hline
\ Day & Number of nodes & Number of edges & Density \\
\hline
1 & 213 & 1009 & 0.045 \\
2 & 212 & 1099 & 0.049 \\
3 & 219 & 1249 & 0.052 \\
4 & 208 & 1108 & 0.051 \\
5 & 209 & 992 & 0.046 \\
6 & 208 & 975 & 0.045 \\
7 & 210 & 1004 & 0.046 \\
8 & 210 & 976 & 0.044 \\
9 & 210 & 1106 & 0.050 \\
10 & 209 & 1006 & 0.046 \\
11 & 205 & 980 & 0.047 \\
12 & 210 & 1045 & 0.048 \\
13 & 209 & 1046 & 0.048 \\
14 & 205 & 1095 & 0.052 \\
\hline
\end{tabular}
\caption{ Statistics for the graph of each day }
\label{tab:graph_stat}
\end{table}

Numbers of nodes, numbers of edges, and the densities of these networks are relatively stable within fourteen days. The densities of these graphs show that the adjacency matrices corresponding to these graphs are rather sparse. This indicates that there may be few connections between the cows. Furthermore, the connections between the cows are loose, or some connections may be missing.

The 21 nodes (about 10\% of the total cows) with the highest degree in the graph of each day are retrieved. Then we count the nodes that appeared in the list of nodes with the highest degree more than seven times in 14 days. The results are shown in the Table \ref{tab:times}. The first three nodes in the table could be considered as the most critical nodes in the social network. The three cows match the result in Figure \ref{fig:centrality} (a), where there are three nodes with the highest centrality.
\begin{table}[ht!]
\centering
\begin{tabular}{|c|c|}
\hline
\ Cow ID & Times  \\
\hline
2202988 & 11 \\
2203574 & 10 \\
2202720 & 10 \\
\hline
\end{tabular}
\caption{ Nodes that appear more than 7 times }
\label{tab:times}
\end{table}
\subsubsection{Similarity of cows' daily social networks}
The problem of comparing social network similarity can be viewed as a graph matching problem, which is an NP-hard problem \cite{caetano2009learning}. Therefore, we compare the corresponding adjacency matrices' eigenvectors to investigate the similarity \cite{SHEMESH198411}. The similarity metric is then the sum of the squared differences between the largest k eigenvalues between the graphs. This produces a similarity metric in the range $[0,+\infty)$, where values closer to zero are more similar.
\begin{figure}[ht!]
 \centering
 \includegraphics[width=120mm]{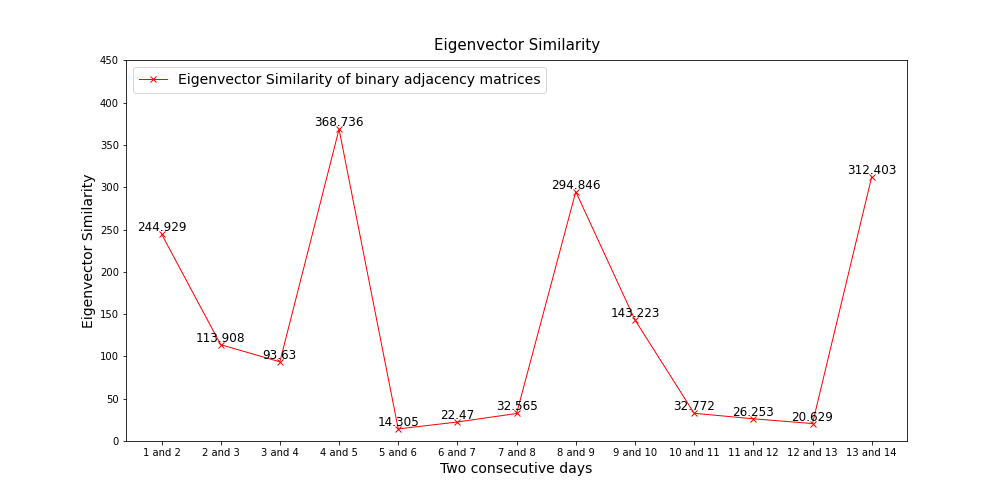}
 \caption{Eigenvector Similarity of Binary Adjacency Matrices}
 \label{fig:graphsimilarity}
\end{figure}

The results are computed based on the binary adjacency matrices. Most day-long association matrices are significantly positively correlated, while some are relatively independent. The values between consecutive matrices range from 14.305 to 368.736.

\subsection{Results of community detection}
\subsubsection{Communities of the whole barn area}
Different measures of community detection are used when choosing the ideal community detection method.

Modularity is a measure of the structure of networks that measures the strength of division of a network into modules (also called groups, clusters, or communities). The communities should not be overlapping. The minimum value of modularity is -0.5, and the maximum value is 1.
Greater modularity means a better community partition.
When the network is not divided into communities, i.e., the nodes in the network are in the same community, the value of modularity is 0.
Newman pointed out that in practice, the value of modularity of a good partition should be larger than 0.3 \cite{newmanarticle}.
The modularity of partitions generated by different methods is measured and shown in the Table \ref{tab:mod}. The modularity does not apply to CPM because the communities generated by CPM may be overlapping.

\begin{table}[ht!]
\centering
\begin{tabular}{|c|c|c|c|}
\hline
\ Day & Girvan-Newman & Louvain(unweighted) & Louvain(weighted) \\
\hline
1 & 0.030 &0.327&0.488 \\
2 & 0.014 &0.296&0.446 \\
3 & 0.031 &0.302&0.470 \\
4 & 0.013 &0.287&0.457 \\
5 & 0.054 &0.329&0.486 \\
6 & 0.050 &0.317&0.493 \\
7 & 0.059 &0.322&0.488 \\
8 & 0.034 &0.330&0.490 \\
9 & 0.056 &0.316&0.492 \\
10 & 0.026 &0.321&0.478 \\
11 & 0.011 &0.324&0.489 \\
12 & 0.027 &0.300&0.483 \\
13 & 0.044 &0.320&0.476 \\
14 & 0.015 &0.302&0.451 \\
\hline
\end{tabular}
\caption{ Modularity of partitions generated by different algorithms }
\label{tab:mod}
\end{table}
The number of communities is set in advance when applying GN on the unweighted graphs. Different values have been tested. Unfortunately, none of them resulted in good partitions. The number of communities is set to be 20 for the presented results. On the network data of one day, the sizes of the communities generated by the algorithm vary greatly, and there are large communities with more than 150 nodes. This indicates that the algorithm has almost no clustering effect on more than 70\% of the nodes in the network. As seen in Table \ref{tab:mod}, the modularity is quite bad. Thus the GN algorithm is not suitable for this application.

\begin{figure}[ht!]
 \centering
 \includegraphics[width=120mm]{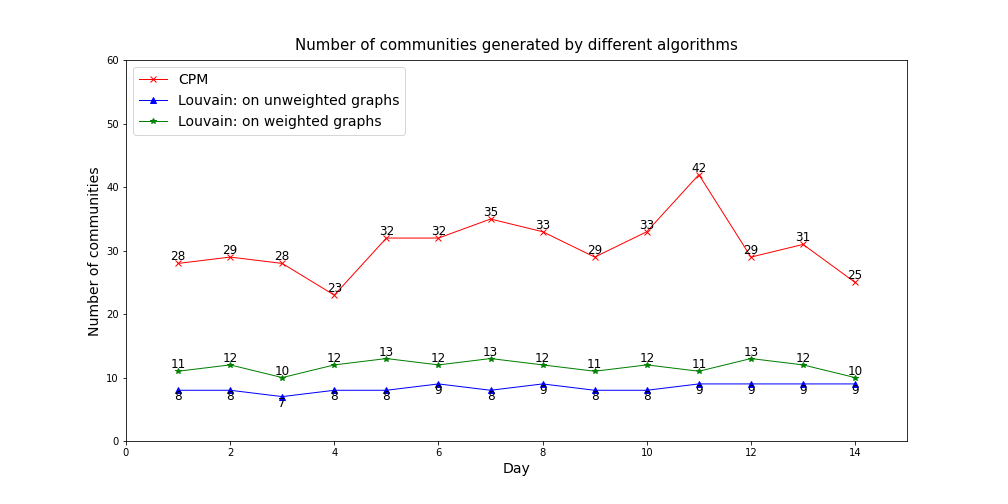}
 \caption{Number of communities generated by different algorithms}
 \label{fig:numofcom}
\end{figure}

Because the data comes from fourteen consecutive days, it is assumed that the number of communities should not change significantly. 
Figure \ref{fig:numofcom} shows the numbers of communities generated by different methods over the 14 days. The number of communities is more stable for the Louvain method than CPM.

\begin{figure}[ht!]
 \centering
 \includegraphics[width=120mm]{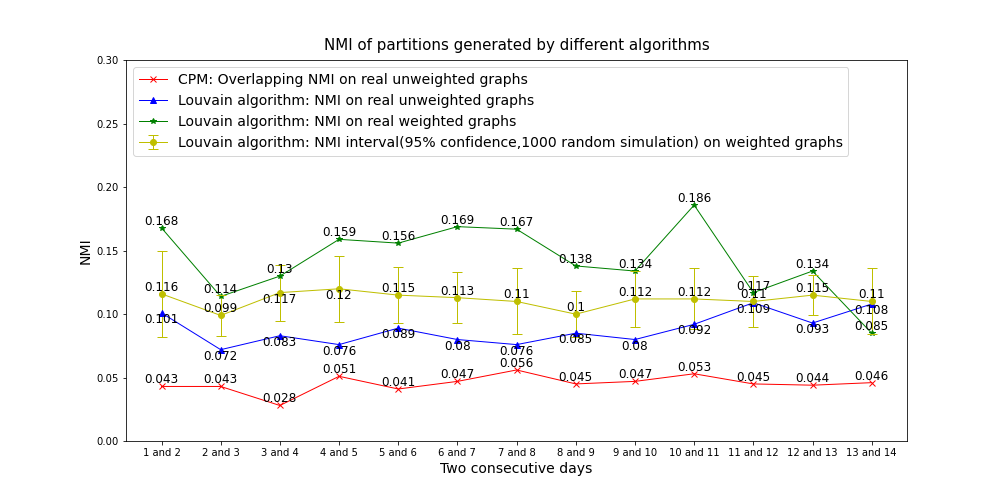}
 \caption{NMI of partitions generated by different algorithms on different graphs}
 \label{fig:nmi}
\end{figure}

Normalized Mutual Information (NMI) is a measure used to evaluate network partitioning performed by community-finding algorithms \cite{yang2016comparative}. It is often considered due to its comprehensive meaning and its ability to allow the comparison of two partitions even when they have a different number of clusters. The closer the NMI is to one, the more consistent information between the two partitions is. The closer the NMI is to zero, the less consistent the information between the two partitions is. For example, if the partition of day 1 is the same as day two, the NMI of day one and day two will be one; if the partitions are totally different from each other, the NMI will be zero. Usually, NMI is only used in those communities which are not overlap. To measure the partition of CPM, a modified version of NMI called overlapping NMI is introduced \cite{McDaidNMI}. Figure \ref{fig:nmi} shows the NMI between two consecutive days.

We see in Figure \ref{fig:numofcom} that for CPM, the number of generated communities from the unweighted graph has significantly changed in these fourteen days, which is different from the assumed conditions. And the overlapping NMI of CPM is close to zero, which means the division of communities each day is quite independent. 

From Table \ref{tab:mod} and Figure \ref{fig:numofcom}, it is found that the Louvain algorithm has a relatively good result. The number of communities is stable. The modularity is relatively high. Compared with the unweighted version, the weighted version of Louvain has better partitions. 

The NMI of Louvain is also measured as shown in Figure \ref{fig:nmi}. The weighted version of Louvain has a higher NMI than the unweighted version. In addition, random weighted graphs with the same density as real weighted graphs are created to simulate the social network. For instance, a random weighted graph is created with the density of the graph of day 1. Then weighted Louvain algorithm is applied to the random graph, and the partition of the random graph is generated. The NMI was calculated using the partitions of random graphs. We performed 1000 random simulations. The mean and 95\% confidence interval of the NMI from the randomly generated data is shown with a yellow line in Figure \ref{fig:nmi}. 

In most of the fourteen days, NMI on real weighted graphs is higher than on random graphs. This shows that in most cases, there are certain communities structured in the cows' social network. However, the highest NMI is still less than 0.2, which means a big difference between the partitions of two consecutive days. To a large extent, cow communities are relatively independent in time.

\subsubsection{Communities of four specific areas}
\begin{figure}[ht!]
 \centering
 \includegraphics[width=120mm]{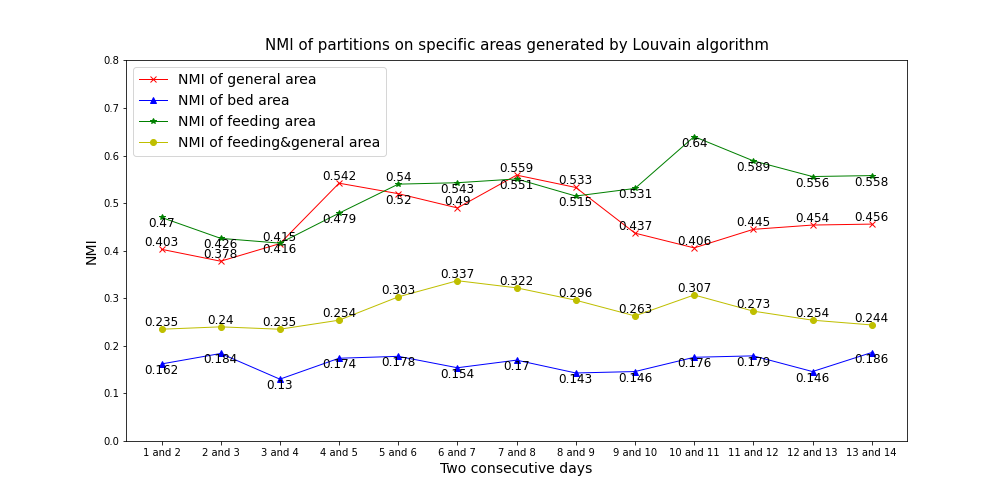}
 \caption{NMI of partitions generated by different algorithms on different graphs}
 \label{fig:nmi2}
\end{figure}

As mentioned in Section 4.6.2, the cows are considered to prefer different companions for different activities. Thus we apply Louvain algorithm on weighted graphs of the four specific areas as shown in Figure \ref{fig:barn}. Also, we combine the data of feeding area and general area to get new adjacency matrices. The corresponding modularity is shown in the Table \ref{tab:mod2}.
\begin{table}[ht!]
\centering
\begin{tabular}{|c|c|c|c|c|}
\hline
\ Day & Feeding area & Bed area & General area & Feeding\&general area \\
\hline
1 & 0.775 &0.597&0.726 &0.687 \\
2 & 0.831 &0.519&0.776 &0.737 \\
3 & 0.765 &0.577&0.776 &0.709 \\
4 & 0.887 &0.538&0.794 &0.757 \\
5 & 0.800 &0.555&0.794 &0.753 \\
6 & 0.879 &0.561&0.819 &0.769 \\
7 & 0.798 &0.569&0.847 &0.799 \\
8 & 0.867 &0.554&0.807 &0.777 \\
9 & 0.719 &0.584&0.785 &0.723 \\
10 & 0.879 &0.553&0.759 &0.744 \\
11 & 0.888 &0.590&0.812 &0.763 \\
12 & 0.751 &0.559&0.768 &0.713 \\
13 & 0.863 &0.559&0.771 &0.744 \\
14 & 0.804 &0.514&0.730 &0.706 \\
\hline
\end{tabular}
\caption{ Modularity of partitions on specific areas generated by weighted Louvain algorithm }
\label{tab:mod2}
\end{table}
In addition, the corresponding NMI is shown in Figure \ref{fig:nmi2}.

The robot area is where cows get milked. The threshold of 30 minutes is too long to detect communities. Thus in this area, there is no generated community. Some other thresholds can be test in the future.

The modularity and NMI improve a lot in the feeding area and general area. These results show that the communities in these areas are relatively stable over the 14 days. The modularity of the bed area also improves while the NMI of it is worse than when looking at the whole barn. This demonstrates that sleeping is an activity with less social meaning. The results of the combination of the general area and feeding area illustrate the combination of two good results does not lead to better results but leads to worse results. We cannot simply combine different activities of cows to detect communities.

The value of NMI that yields very stable communities is unknown and more research is needed \cite{Zhang_2015}. Therefore, we checked whether an NMI of about 0.55 is large enough to represent the generation of stable communities. We investigated the communities over these days in general area and feeding area manually. In the communities of three consecutive days, the communities of the first two days have a maximum of four nodes that overlap. But these four nodes are not in the same community on the third day. According to the results of manual inspection, the communities are still not stable enough.

\subsubsection{Visualization of communities and social network}
\begin{figure}[H]
 \centering
 \includegraphics[width=150mm]{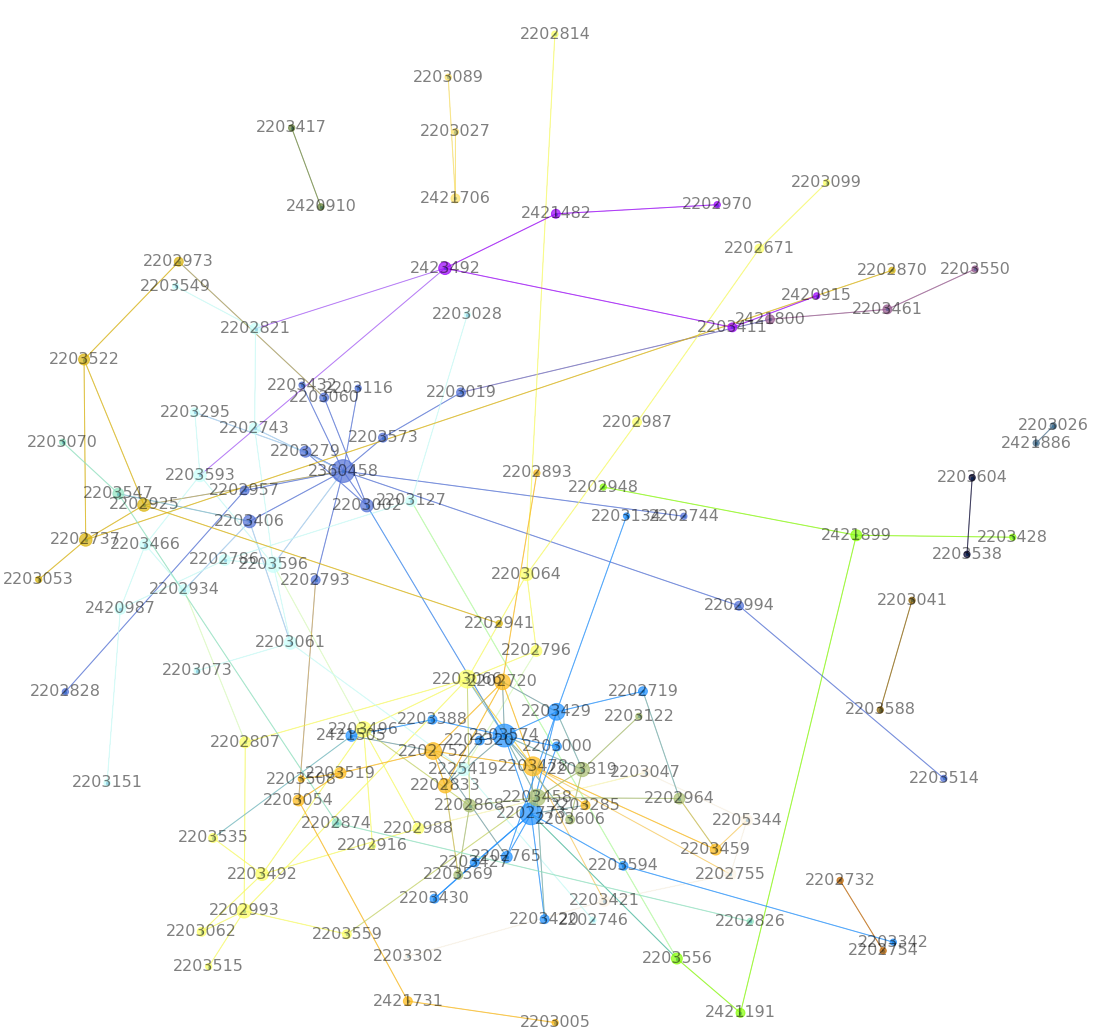}
 \caption{Visualization of communities of Day 1 using Louvain based on weighted graphs of the general area}
 \label{fig:vis}
\end{figure}
For better analysis, the social networks and communities are visualized. Figure \ref{fig:vis} shows an example of detected communities. Each node in the graph represents a cow. The edges between the cows represent the relationship between the cows. The size of the node is positively related to the degree of the node. The colors of nodes and edges are determined by the communities they belong to. The layouts of graphs are computed automatically by the package {\tt networkX}.

\section{Conclusions}
The centrality distribution can be seen as a distribution of how important the cow is in the network, where there seems to exist some social hierarchy. According to the centrality distribution and nodes degree, we can draw the conclusion that this hierarchy is not very dynamic since at least the cows at the top of the hierarchy are pretty constant. In addition, we see that there is a significant chance that the edge distributions from different days are coming from the same distribution, and thus are correlated.

Our results in the monadic measurements clearly show that there is a monadic social structure that is constant enough over the 14 days to be seen as significantly constant. The social structure could be seen as some sort of hierarchy. This is a similar conclusion as in the previous study \cite{ROCHA2020104921}.


The transitivity shows that our network does not cluster, but the network is still closely connected according to the average shortest path. This means that the cows are making connections, but that a connection between two cows does not influence the other connections it has. Our results show that the network characteristics are similar between days with minor differences.

The conclusions from the eigenvalues and eigenvectors are similar to those from the monadic measurements above. Eigenvalues and eigenvectors can be used to show the eigenvector centrality, which can be used instead of the degree centrality, or the spectral radius which approximates the average degree distribution. It can also do more things like transitivity \cite{rohrich_2021} and partitioning of the network into two different groups using spectral graph partitioning \cite{slininger} . The advantage of using the eigenvalues is that it is faster to compute compared with the monadic measurements, and we only have to compute it once to get all the other characteristics.



The undesirable results of CPM can be partially explained from the technical side. CPM works well on dense graphs but poorly on sparse graphs. With the current threshold, the social networks are sparse, which means that CPM did not perform well.

The weighted Louvain algorithm is the most promising community detection algorithm of all the ones we used. We see that there is a community structure in the cow's social network; it is, however, unstable. This is similar to the findings of previous studies \cite{BOYLAND20161}. The number of communities generated every day is relatively stable. However, the NMI obtained by the partitions of the neighboring two days of the community is very low. The nodes included in the community are changing drastically every day. 

When partitioning the barn into areas, the NMI for the feeding and general is drastically higher than with the non-partitioned barn. We conclude that meaningful social interactions only happen in these areas. We see that when combining the areas, the NMI decreases, which means that the social connections are different in the different areas. We also see that the bed area acts as noise and should be disregarded in further research. No conclusions could be reached on the milking area(robot area), since the time spent there is generally lower than the time threshold for the considered connections. Other thresholds need to be studied.

In general, we can conclude that the considered networks did not indicate any stable community structure at the barn level; however, we found significant centralization, which is consistent over the whole period. Relationships between individuals were differentiated, with cows associating non-randomly. This study demonstrates the possibility of detecting relatively stable communities in specific areas and social network analysis for understanding social relationships in cow groups, both of which are likely to play an important role in future research into social networks of dairy cows, and more generally, the social networks of livestock.

\section{Future improvements}
The time limitations for this project impose constraints on what could be studied and achieved. We believe numerous improvements could give a better or different insight into dairy cows' dynamic social network.
Investigating the use of area-specific matrices is the most promising approach to find communities, and we believe developing this approach further has the highest chance of yielding good results.
Weighted graphs may improve the accuracy of community detection. Thus, more community detection methods applied to weighted graphs could be explored, especially those methods that can generate overlapping communities that match the characteristics of cows' social networks. 
The adjacency matrices of graphs in the project are sparse. The sparsity affects the result of community detection methods. We recommend that the time thresholds used to create adjacency matrices be decreased to see if any improvement appears.
Another promising direction is the integration of data from multiple farms to build more generalized and robust models of cows' social interactions. Due to privacy, ownership, and logistical constraints, directly sharing raw data across farms is often not feasible. 
Federated learning (FL) and distributed learning (DL) offer practical solutions to this problem by allowing model training across decentralized datasets while preserving data locality~\cite{yang2019federated, li2020federated, chu2023resource, konevcny2016federated, mcmahan2017communication}. These paradigms can enhance the scalability and adaptability of network modeling techniques by leveraging heterogeneous data sources.

In addition, the use of large language models (LLMs) for behavioral pattern recognition and anomaly detection has shown considerable promise. 
LLMs can capture complex temporal and contextual patterns from structured and unstructured data, including sensor streams and metadata~\cite{bommasani2021opportunities, zhang2023llm4agri, liu2023summary}. 
Applying LLMs to model cows' interactions and predict social behavior could lead to more accurate insights into social structures. 
However, one of the critical challenges of LLM adoption is the risk of privacy leakage through model inversion or training/inference data extraction attacks~\cite{carlini2021extracting, lehman2021does, chu2024reconstruct, zhang2021understanding}. 
Despite this, LLMs remain among the most promising tools for complex behavioral modeling due to their generalization capability, scalability, and zero-shot learning potential~\cite{brown2020language, touvron2023llama, openai2023gpt4}.

\newpage

\bibliographystyle{ieeetr} 
\bibliography{references} 

\appendix
\end{document}